\documentclass[prd,aps,twocolumn,preprintnumbers,amsmath,amssymb,nofootinbib,superscriptaddress]{revtex4}
\usepackage{graphicx}
\usepackage{overpic}
\usepackage[utf8]{inputenc}
\usepackage{color}
\usepackage{epstopdf}
\usepackage[colorlinks=true,
linkcolor=blue,
breaklinks=true,
urlcolor=blue,
citecolor=blue]{hyperref}

\begin{document}

\title{Light baryon spectroscopy at BESIII}

\author{Ronggang Ping}
\email{pingrg@ihep.ac.cn}
\affiliation{Institute of High Energy Physics, Chinese Academy of Sciences, Beijing, 100049, China}

% 摘要必须在 \maketitle 之前
\begin{abstract}
Based on the data sample of $2.7\times 10^{9}$ $\psi(3686)$ events collected with the BESIII detector, this contribution presents recent advances in baryon spectroscopy. Key results from amplitude analyses of the decay channels $\psi(3686) \to p\bar{p}\pi^{0}$, $\psi(3686) \to p\bar{p}\eta$, and $\psi(3686) \to \Lambda\bar{\Sigma}^{0}\pi^{0} + c.c.$ are summarized. Prospects for future baryon spectroscopy studies at BESIII are also discussed.
\end{abstract}


\maketitle

\section{Introduction}

The nature of exotic baryonic resonances continues to challenge conventional quark models, revealing complex structures beyond three-quark configurations. The $N(1535)$ resonance exemplifies this, with experimental measurements showing significant uncertainties and an unusually large branching fraction ratio $\Gamma_{N\eta}/\Gamma_{N\pi} \approx 1.0\pm0.4$ \cite{pdg}. This strong coupling to the strange $\eta N$ channel, combined with the ``mass reverse" problem where its mass exceeds that of the radial excitation $N(1440)$, suggests substantial exotic components. Theoretical interpretations primarily focus on two scenarios: a dynamically generated molecular state from $S$-wave meson-baryon interactions within the chiral unitary approach \cite{molecu1,molecu2}, and a pentaquark picture with $uuds\bar{s}$ configuration \cite{penta1,penta2,penta3,penta4}. Recent analyses of $\Lambda_c^+ \to p\bar{K}^0\eta$ decays by the Belle collaboration favor the molecular description, indicating a complex structure beyond the pure three-quark state \cite{Belle}.

Similarly, the $\Lambda(1405)$ resonance with strangeness $S=-1$ and isospin $I=0$ represents a paradigm of dynamically generated states. Studies within the chiral unitary approach reveal that $\Lambda(1405)$ possesses a distinctive two-pole structure in the complex energy plane, arising from attractive interactions in both the singlet and octet channels of SU(3) flavor symmetry \cite{chi1,chi2,chi3,chi4}. The higher pole is located near the $\bar{K}N$ threshold at approximately $(1420 - i20)$ MeV, while the lower pole appears closer to the $\pi\Sigma$ threshold around $(1380 - i80)$ MeV. This dual nature explains the reaction-dependent lineshapes observed experimentally. Precision measurements by the SIDDHARTA collaboration \cite{bazzi1,bazzi2} and recent analyses incorporating next-to-leading order chiral interactions continue to refine our understanding of this fundamental resonance.

In contrast, the $\Lambda(2325)$ with possible spin-parity $J^P = 3/2^-$ remains less explored. Currently omitted from the PDG summary table, its experimental evidence primarily comes from early partial-wave analyses of $K^- p \to \Lambda\omega$ and $K^- p \to \overline{K} N$ reactions. BACCARI \cite{deBellefon:1975yp} and DEBELLEFON \cite{deBellefon:1976qr} reported a mass of approximately $2325$ MeV and a width of $160$--$177$ MeV. Although partial-wave analyses favor $3/2^-$, the possibility of $3/2^+$ cannot be excluded, and quantum number assignments still rely on model-dependent assumptions. With $NK$ as the dominant decay mode (branching fraction $0.19\pm 0.06$) and limited theoretical studies available, further experimental and theoretical efforts are crucial to confirm its properties and internal structure.

\section{Experimental Study of $N(1535)$ in $\psi(3686) \to p\bar{p}\pi^0$ and $p\bar{p}\eta$ Decays}

Using a sample of $(2712 \pm 14) \times 10^6$ $\psi(3686)$ events collected with the BESIII detector, partial wave analyses (PWA) of the decays $\psi(3686) \to p\bar{p}\pi^0$ and $\psi(3686) \to p\bar{p}\eta$ have been performed \cite{BESIII:2024vqu}. After event selection, a total of 190,729 candidates for $\psi(3686) \to p\bar{p}\pi^0$ and 31,441 candidates for $\psi(3686) \to p\bar{p}\eta$ are retained. The Dalitz plots for both channels exhibit clear structures indicating the presence of excited $N^*$ resonances and $p\bar{p}$ intermediate states, providing a rich basis for amplitude analysis.

\begin{figure}[htbp]
\centering
\begin{minipage}{0.48\textwidth}
\centering
\begin{overpic}[width=\linewidth]{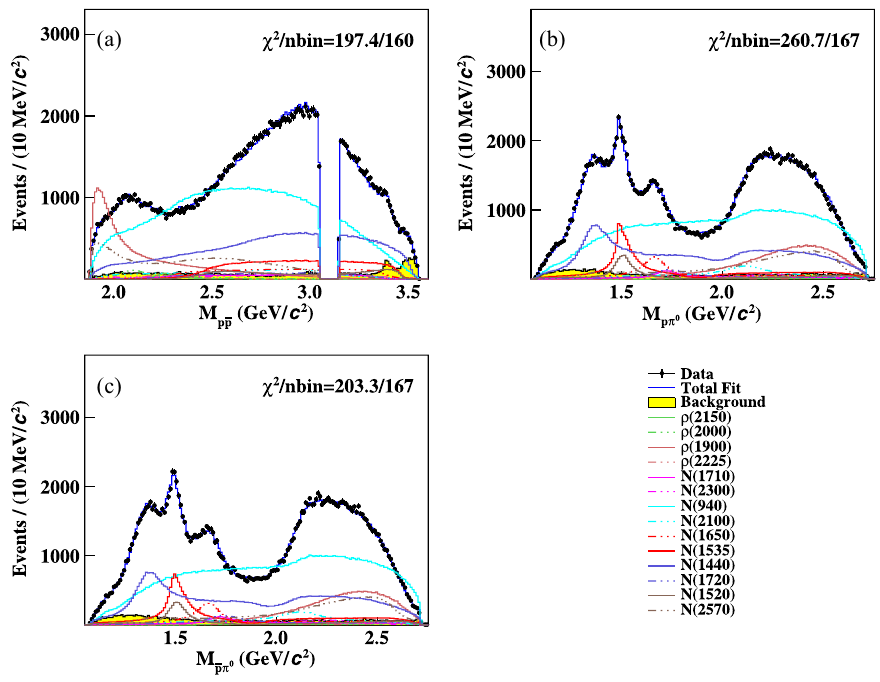}
\end{overpic}
\end{minipage}
\hfill
\begin{minipage}{0.48\textwidth}
\centering
\begin{overpic}[width=\linewidth]{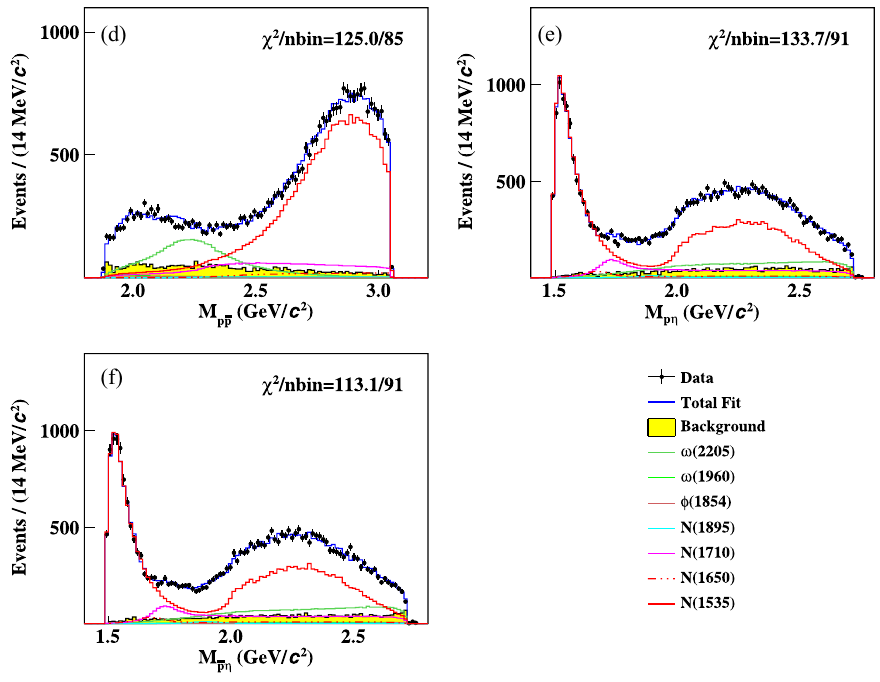}
\end{overpic}
\end{minipage}
\caption{Projections of the PWA fits for $\psi(3686) \to p\bar{p}\pi^0$ on (a) $M_{p\bar{p}}$, (b) $M_{p\pi^0}$, (c) $M_{\bar{p}\pi^0}$ and for $\psi(3686) \to p\bar{p}\eta$ on (d) $M_{p\bar{p}}$, (e) $M_{p\eta}$, (f) $M_{\bar{p}\eta}$.}
\label{fig:ppbarpi}
\end{figure}

The PWA incorporates well-established $N^*$ states with $J \leq 5/2$ and significant couplings to $N\pi$ or $N\eta$, including $N(1440)$, $N(1520)$, $N(1535)$, $N(1650)$, $N(1710)$, $N(1720)$, $N(1895)$, $N(2100)$, $N(2300)$, and $N(2570)$, as well as virtual proton pole contributions. The $p\bar{p}$ systems are described using excited $\rho$, $\omega$, and $\phi$ mesons such as $\rho(1900)$, $\rho(2000)$, $\rho(2150)$, $\rho(2225)$, $\phi_3(1850)$, $\omega(1960)$, and $\omega(2205)$. The baseline solutions provide a good description of the data, with all included resonances having statistical significances greater than $5\sigma$. Fit results are shown in Fig. \ref{fig:ppbarpi}.

Branching fractions for several $N^*$ intermediate states are also measured. The ratio of branching fractions $\mathcal{B}(\psi(3686)$ $ \to p\bar{p}\pi^0) / \mathcal{B}(J/\psi \to p\bar{p}\pi^0)$ is found to be approximately $(11.3$-$15.4)\%$, consistent with the ``12\% rule" for charmonium decays, while the corresponding ratio for $p\bar{p}\eta$ is about $(3.1$-$4.2)\%$, indicating a significant violation in the $\eta$ channel. This work confirms the exotic nature of $N(1535)$ in $e^+e^-$ collisions for the first time and provides crucial input for understanding its structure beyond the conventional three-quark model.

To accurately estimate the continuum background contribution in the $\psi(3686)$ resonance region, a PWA is performed using a data sample collected at $\sqrt{s} = 3.773\ \mathrm{GeV}$ with an integrated luminosity of $2.93\ \mathrm{fb}^{-1}$. This energy point lies outside the $\psi(3686)$ resonance peak and serves as a control sample to model the non-resonant $e^+e^-$ annihilation background. The analysis assumes that the observed yields at this energy are dominated by the continuum process, while potential contributions from $\psi(3770) \to p\bar{p}\pi^0/\eta$ decays are explicitly ignored, as no significant evidence for such decays has been observed with the current statistics.

For consistency and to facilitate a direct comparison and subsequent background subtraction, the PWA at $\sqrt{s} = 3.773\ \mathrm{GeV}$ employs the same set of intermediate states established in the baseline solution at $\sqrt{s} = 3.686\ \mathrm{GeV}$. This includes the well-established $N^*$ resonances such as $N(1440)$, $N(1520)$, $N(1535)$, $N(1650)$, and higher states, as well as the $p\bar{p}$ mesonic states like $\rho(1900)$, $\rho(2000)$, and $\omega(1960)$. The fitted yields for these intermediate states at $\sqrt{s} = 3.773\ \mathrm{GeV}$ are then normalized and subtracted from the corresponding yields obtained at the $\psi(3686)$ peak. This procedure allows for the extraction of the net signal yields originating purely from the $\psi(3686)$ decay, thereby providing a cleaner measurement of its branching fractions and the properties of the $N^*$ resonances produced therein.

A key result of this analysis is the precise measurement of the decay width ratio:
\[
\frac{\Gamma_{N(1535) \to N\eta}}{\Gamma_{N(1535) \to N\pi}} = 0.99 \pm 0.05_{\text{stat.}} \pm 0.19_{\text{syst.}}.
\]
This value is consistent with the PDG average of $1.00 \pm 0.40$ from fixed-target experiments, but with significantly improved precision. The large coupling to $N\eta$ suggests a substantial $s\bar{s}$ component in the $N(1535)$ wave function, supporting exotic structure interpretations such as a $uuds\bar{s}$ pentaquark configuration or a dynamically generated state with strong strange quark content. Further confirmation in the $K\Lambda$ final state is needed to clarify its internal composition.

\section{Partial Wave Analysis of $\psi(3686)\to\Lambda\bar{\Sigma}^{0}\pi^{0}$}

Based on a sample of $(2712.4\pm14.3)\times10^{6}$ $\psi(3686)$ events collected with the BESIII detector, a comprehensive PWA of the decay $\psi(3686)\to\Lambda\bar{\Sigma}^{0}\pi^{0} + \text{c.c.}$ is performed to investigate the excited hyperon spectrum \cite{BESIII:2024jgy}. The decay chain is reconstructed through the sequential processes $\Sigma^{0}\to\gamma\Lambda$, $\Lambda\to p\pi^{-}$, and $\pi^{0}\to\gamma\gamma$, resulting in a final state of $p\bar{p}\pi^{+}\pi^{-}\gamma\gamma\gamma$. After applying stringent event selection criteria, including secondary vertex reconstruction, kinematic fitting with a four-constraint fit, and particle identification, a total of 14,414 candidate events are selected for the analysis. The continuum background from non-resonant $e^{+}e^{-}\to\Lambda\bar{\Sigma}^{0}\pi^{0}$ production at $\sqrt{s}=3.773\ \mathrm{GeV}$ is carefully estimated and normalized, contributing $235.2\pm11.3$ events, which are subsequently subtracted within the negative log-likelihood framework.

The analysis employs the helicity amplitude formalism in conjunction with the isobar model, where the three-body decay is treated as a sequence of two-body processes. The full decay amplitude is constructed using relativistic covariant tensor expressions, incorporating Blatt-Weisskopf barrier factors with a radius parameter of $d=0.73\ \mathrm{fm}$. Background contributions from mis-combination, $\pi^{0}$ sidebands, and continuum processes are explicitly incorporated and subtracted in the unbinned maximum likelihood fit. The nominal fit model includes established $\Lambda^{*}$ and $\Sigma^{*}$ resonances with at least three-star status in the PDG and spins below $5/2$. All resonance masses and widths are treated as free parameters in the fit to ensure minimal model dependence.

\begin{figure}[htbp]
\centering
{
\begin{overpic}[width=0.6\textwidth]{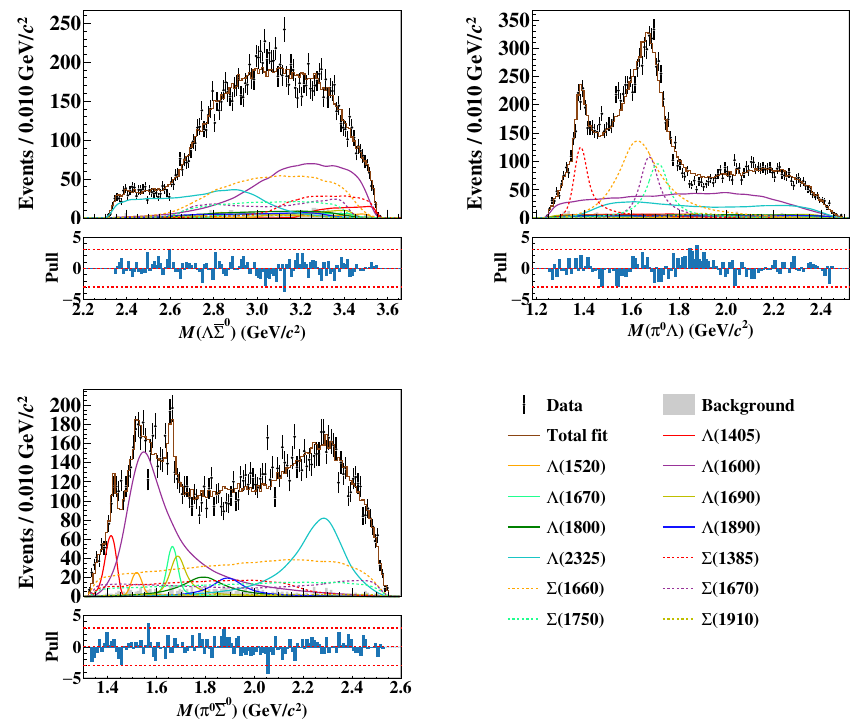}
\put(15,57){(a)}
\end{overpic}
}
\caption{Projections of the PWA fits for $\psi(3686) \to \Lambda\bar{\Sigma}\pi^0$ on  $M_{\Lambda\bar\Sigma^0}$, $M_{\pi^0\Lambda}$, and $M_{\pi^0\bar\Sigma^0}$.}
\label{fig:lambda}
\end{figure}

As shown in Fig. \ref{fig:lambda}, the invariant mass distribution of $\Lambda\bar{\Sigma}^{0}$ is successfully described without introducing $\rho^{*}$ or $a^{*}$ mesonic states, while the $M(\pi^{0}\Lambda)$ spectrum is modeled using $\Sigma^{*}$ resonances below $2.0\ \mathrm{GeV}$. A particularly significant finding is the necessity of including the $\Lambda(2325)$ state to describe the high-mass region of the $M(\pi^{0}\bar{\Sigma}^{0})$ spectrum. This resonance, previously listed with one-star status in the PDG, emerges as a dominant component with a fit fraction of $(18.9\pm1.3)\%$. Its mass and width are precisely measured to be $M=2306.5\pm6.3\pm17.1\ \mathrm{MeV}/c^{2}$ and $\Gamma=227.1\pm12.2\pm47.8\ \mathrm{MeV}$, respectively. A detailed spin-parity analysis comparing different quantum number hypotheses strongly favors $J^{P}=3/2^{-}$, consistent with previous experimental indications.

For the $\Lambda(1405)$ resonance, the analysis implements a sophisticated chiral unitary approach that properly accounts for its well-established two-pole structure. The propagator takes the form $R(m)=[I-VG]^{-1}$, where $V$ represents the interaction kernel and $G$ the meson-baryon loop function, following the formulation in Ref. \cite{oset}. This parameterization yields a statistical significance of $11.1\sigma$ and a fit fraction of $(3.0\pm0.3)\%$. An alternative fit using a Flatte-like formula, which explicitly couples the $\pi\Sigma$ and $\bar{K}N$ channels, provides a comparable description with a significance of $11.2\sigma$ and a fit fraction of $(6.7\pm0.9)\%$. Both parameterizations adequately describe the data, though the chiral dynamics approach is preferred in the nominal fit due to its stronger theoretical foundation in QCD chiral symmetry.

The systematic uncertainties are thoroughly investigated, including contributions from background estimation, resonance parameterizations, barrier factor radius variations, and potential additional resonance components. The total systematic uncertainties for the $\Lambda(2325)$ parameters are found to be 17.1 MeV/$c^{2}$ for the mass and 47.8 MeV for the width, dominated by uncertainties from extra resonance considerations and background modeling.

This analysis represents the first amplitude-based study of $\psi(3686)\to\Lambda\bar{\Sigma}^{0}\pi^{0}$, providing crucial evidence for the existence of the $\Lambda(2325)$ and reinforcing the two-pole nature of the $\Lambda(1405)$. The results offer valuable inputs for understanding the spectrum and internal structure of strange baryons, and demonstrate the capability of $e^{+}e^{-}$ colliders to probe exotic baryonic states beyond the conventional three-quark model.

\section{Summary and Outlook}

Recent spectroscopic studies at BESIII have unveiled exotic structures in light baryons. The $N(1535)$ resonance exhibits a confirmed large coupling to the $N\eta$ channel, with $\Gamma_{N\eta}/\Gamma_{N\pi} = 0.99 \pm 0.05_{\text{stat.}} \pm 0.19_{\text{syst.}}$, providing strong evidence for a significant $\bar{s}s$ component. Furthermore, the observation of the $\Lambda(1380)$ ($>11\sigma$) in $\psi(3686) \to \Lambda\bar{\Sigma}^0\pi^0$ decays offers crucial support for the two-pole structure of the $\Lambda(1405)$, while a new $\Lambda(2325)$ state with preferred quantum numbers $J^P=3/2^-$ has been established. These systematic investigations significantly advance our understanding of baryon structure and exotic generation mechanisms, paving the way for future explorations into non-perturbative QCD.

\end{document}